\documentclass[proof]{pasj01}
\Received{$\langle$reception date$\rangle$}
\Accepted{$\langle$acception date$\rangle$}
\Published{$\langle$publication date$\rangle$}

\begin{document}

\title{Constraints on Pulsed Emission Model for Repeating FRB 121102}
\author{Shota \textsc{Kisaka}  \altaffilmark{1,*},
 Teruaki \textsc{Enoto} \altaffilmark{2,3},
  and Shinpei \textsc{Shibata} \altaffilmark{4}}%
\altaffiltext{1}{Department of Physics and Mathematics, Aoyama Gakuin University, Sagamihara, Kanagawa, 252-5258, Japan}
\altaffiltext{2}{The Hakubi Center for Advanced Research, Kyoto University, Kyoto 606-8302, Japan}
\altaffiltext{3}{Department of Astronomy, Kyoto University, Kitashirakawa-Oiwakecho, Sakyo-ku, Kyoto, 606-8502, Japan}
\altaffiltext{4}{Department of Physics, Yamagata University, Kojirakawa, Yamagata, 990-8560, Japan}

\email{kisaka@phys.aoyama.ac.jp}


\maketitle

\begin{abstract}
Recent localization of the repeating Fast Radio Burst (FRB) 121102 revealed 
the distance of its host galaxy and luminosities of the bursts. 
We investigated constraints on the young neutron star (NS) model, that 
(a) the FRB intrinsic luminosity is supported by the spin-down energy, and
(b) the FRB duration is shorter than the NS rotation period. 
In the case of a circular cone emission geometry, 
conditions (a) and (b) determine the NS parameters within very small ranges, 
compared with that from only condition (a) discussed in previous works. 
Anisotropy of the pulsed emission does not affect the area of the allowed parameter region
by virtue of condition (b). 
The determined parameters are consistent with those independently limited 
by the properties of the possible persistent radio counterpart
and the circumburst environments such as surrounding materials. 
Since the NS in the allowed parameter region is older than the spin-down timescale, 
the hypothetical GRP-like model expects a rapid radio flux decay of $\lesssim1$ Jy within a few years  
as the spin-down luminosity decreases. 
The continuous monitoring will give a hint of discrimination of the models.
If no flux evolution will be seen, we need to consider an alternative model,
e.g., the magnetically powered flare.

\end{abstract}

\section{INTRODUCTION}
\label{sec:introduction}

Among fast radio bursts (FRBs), which are radio transients 
with duration of milliseconds and large dispersion measure (DM)
compared with expected for propagation through the Galaxy (e.g., \cite{Lor+07,Tho+13}), 
FRB 121102 was identified as a repeating source \citep{Spi+16}. 
In observations over 2012-2016, 
30 bursts were reported from FRB 121102 at $1.1-3.5$ GHz with the same DM$\sim560$ cm$^{-3}$ pc 
\citep{Spi+14, Spi+16, Sch+16, Cha+17, Mar+17}. 
The flux density was $\sim0.02-3.72$ Jy, and no temporal evolution of the flux 
was apparently seen \citep{Mar+17}.
The Gaussian FWHM pulse width was $2.8-8.7$ ms \citep{Spi+16}. 
Since no scattering tail was observed in the pulses, 
the observed width would be the intrinsic width of the emission \citep{Sch+16}.

Owing to interferometric imaging with the Karl G. Jansky Very Large Array (VLA) 
and European VLBI Network (EVN) observations, 
FRB 121102 was recently localized to $\sim100$ mas and $\sim2-4$ mas precisions \citep{Cha+17, Mar+17}. 
The radio observations also detected a persistent radio source 
with an angular separation from the bursts $\lesssim40$ pc \citep{Mar+17}. 
Optical observations identified the host galaxy 
at a redshift $z=0.19273(8)$, corresponding to a luminosity distance of $972$ Mpc \citep{Ten+17}.
The measured distance gives the luminosity of FRB 121102, 
$L_{\rm FRB}\sim(0.03-6)\times10^{42}$ erg s$^{-1}$ \citep{Mar+17}. 

Repetition of a FRB rules out the catastrophic origin. 
Although magnetically-powered flares from a highly magnetized neutron star (NS) have been suggested 
as candidate FRB sources (e.g., \cite{PP07, L14, K16a}), 
significant constraints on FRB-like radio bursts during the giant flare were given by \citet{TKP16}. 
The giant radio pulse (GRP) like emission from a young, energetic pulsar 
have also been suggested as an origin of the repeating FRB 
(e.g., \cite{CW16, K16a, K16b, K16c, K17, LBP16, L17}).
The broad distribution of the spectral index and the pulse width in FRB 121102 at $\sim1.4$ GHz 
has been observed in the GRPs from the Crab pulsar \citep{KSv10, Mikami+16}.

For the rotation-powered model, 
the observed FRB luminosity has to be lower than the spin-down luminosity. 
This constraint gives the allowed range of the NS parameters, the dipole magnetic field 
$\sim 10^{12}-10^{14}$ G and the rotation period $\lesssim 10$ ms (e.g., \cite{L17, MBM17, KM17}), 
although an efficient conversion mechanism to the radio emission is required as discussed in \citet{CW16}. 
If the beam fraction is much lower than unity, the allowed parameter region becomes much large \citep{K16c}.

Different from thermal phenomena, non-thermal emission from pulsars 
should have both on- and off-pulse phases in a rotation period. 
The non-thermal emissions are produced during particle acceleration and creation 
in some limited regions in the magnetosphere such as polar cap (e.g., \cite{DH82}), 
outer gap (e.g., \cite{CHR86}) and current sheet models (e.g., \cite{KSG02}).
Some complex profile structures seen in FRB pulses may reflect the non-uniform distribution of emissivity 
\citep{Spi+16, Ten+17}.
Coherent radio pulsed emission from pulsars such as GRP is also non-thermal emission (e.g., \cite{CW16}).
Hence, in GRP-like emission model, the pulse width of FRBs gives the lower limit on the rotation period. 

In this paper, we consider the GRP-like pulsed emission model for the repeating FRB 121102.
In section 2, we give the constraints on the spin-down luminosity and the rotation period
from the observed luminosity and the pulse width, respectively. 
Two conditions give stringent limit on the allowed parameter range for an NS.
We also discuss prediction for the flux evolution, 
the constraints from the propagation effects, and possible 
NS formation scenarios for the source of FRB 121102 in section 3.

\section{CONSTRAINTS ON GRP-LIKE EMISSION MODEL}
\label{sec:constraint}

\begin{figure*}
  \begin{center}
  \vspace{-5mm}
   \includegraphics[width=120mm, angle=270]{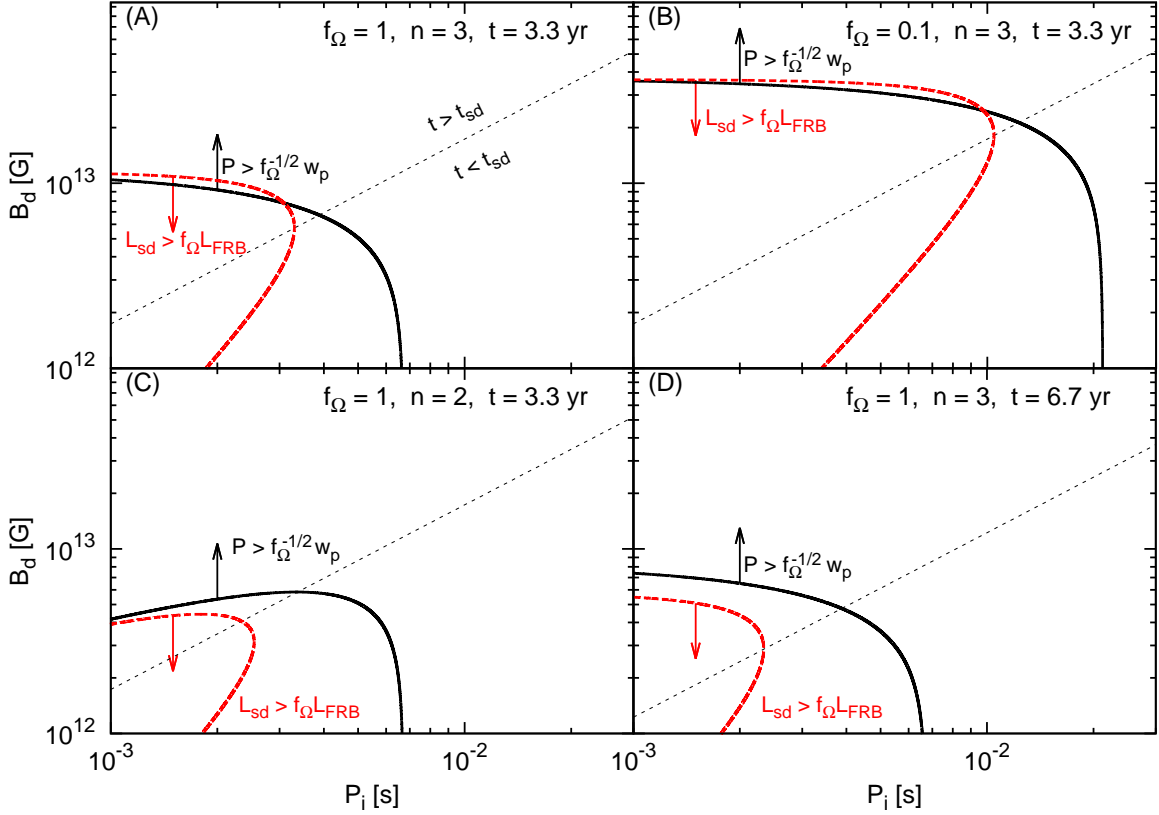}
  \end{center}
   \caption{Constraints on the parameter space $P_{\rm i}$ and $B_{\rm d}$ of 
 GRP-like emission model for FRB 121102. 
 Panel (A); fiducial case with isotropic emission ($f_{\Omega}=1$), 
 dipole spin-down formula ($n=3$) and the lower bound of the age ($t=3.3$ yr). 
 We use the observed values $L_{\rm FRB}=6\times10^{42}$ erg s$^{-1}$ 
 and $w_{\rm p}=8.0~{\rm ms}/(1+z)=6.7$ ms. 
 The red thick dashed lines indicate the luminosity constraints 
 (equation \ref{luminosity_condition}). 
 The black thick solid lines indicate the pulse width constraints 
 (equation \ref{period_condition}). 
 The thin dotted lines indicates the relation $t=t_{\rm sd}$.
 Panel (B); anisotropic emission case ($f_{\Omega}=0.1$). 
 Panel (C); low braking index case ($n=2$). 
 Panel (D); older age of the NS ($t=6.7$ yr). 
 }
   \label{limit}
 \end{figure*}

We consider the GRP-like emission model 
where the emission comes from the pulsar magnetosphere. 
For the emission geometry, we assume a circular cone with angular width 
$\propto\sqrt{f_{\Omega}}$ as widely considered for the radio emission (e.g., \cite{R93}), 
where $f_{\Omega}(\le1)$ is the beaming fraction. 
We use fiducial radius and moment of inertia of an NS at $R_{\rm ns}=12$ km, 
and $I\sim1.4\times10^{45}$ g cm$^2$, respectively. 
Since we focus on a very young NS 
whose age ($\ll10^3$ yr) is much shorter than the decay timescales of dipole magnetic field 
in interior of an NS (e.g., \cite{GR92}),  
we neglect the temporal evolution of the dipole field.

The propagation effects on the radio emission and the possible association 
of persistent radio source could give significant constraints on the model 
for FRB 121102 \citep{MBM17, L17, KM17, CYD17, DWY17}. 
For example, if the injected energy from an NS changes the dynamics of surrounded materials, 
the observed DM could give constraints on the NS parameters. 
This NS nebula model gives the constraints on the NS properties 
as the dipole field $\sim10^{13}-10^{14}$ G, 
the initial rotation period $\sim$ a few ms, 
and the age $\lesssim10^2$ yr \citep{MBM17, L17, KM17, CYD17, DWY17}.
However, the DM is highly sensitive to the unknown ejecta properties such as non-uniformity, 
composition, and ionization state of circumburst materials.  
The detected persistent radio source could be interpreted 
by a low-luminosity AGN \citep{Park+16, Mar+17}, 
so that the persistent source may not be directly related to the FRB source. 
Therefore, we only use the two observed characteristic parameters of the FRB source, 
the isotropic luminosity and the pulse width, 
to constrain on GRP-like emission model in this section.

We consider two conditions to produce the repeating FRB 121102 via GRP-like emission model.
First, the spin-down luminosity $L_{\rm sd}$ has to be larger than the FRB intrinsic luminosity
\footnote{In principle, 
the luminosity of sporadic emissions could exceed the spin-down luminosity if the 
spin-down energy is accumulated and instantaneously released. 
However, unless there are materials with effective inertia
such as supernova fallback disk, (e.g., \cite{M88}), 
the spin-down energy would not be accumulated \citep{L17}. 
Here, we do not consider such materials near the NS.} (e.g., \cite{L17}).  
Taking into account the beaming fraction $f_{\Omega}$ to the intrinsic luminosity, 
the condition on the spin-down luminosity is 
\begin{eqnarray}\label{luminosity_condition}
L_{\rm sd}>f_{\Omega}L_{\rm FRB}, 
\end{eqnarray}
where $L_{\rm FRB}$ is the observed isotropic FRB luminosity.
For the spin-down formula, we assume 
\begin{eqnarray}\label{L_sd}
L_{\rm sd}\sim L_{\rm sd,i}\left(1+\frac{t}{t_{\rm sd}}\right)^{\frac{1+n}{1-n}}, 
\end{eqnarray}
where $t$ is an age of an NS, $n$ is braking index of the rotation period, 
$\dot{P}\propto P^{2-n}$ \citep{MND85}. 
The initial spin-down luminosity $L_{\rm sd, i}$ and the spin-down timescale $t_{\rm sd}$ 
are described by 
\begin{eqnarray}\label{L_sd,i}
L_{\rm sd,i}\sim\frac{B_{\rm d}^2(2\pi/P_{\rm i})^4R_{\rm ns}^6}{2c^3},
\end{eqnarray}
and
\begin{eqnarray}\label{t_sd}
t_{\rm sd}\sim\frac{E_{\rm rot,i}}{L_{\rm sd,i}},
\end{eqnarray}
where $c$ is the speed of light, $P_{\rm i}$ is the initial period, 
$B_{\rm d}$ is the strength of the surface dipole magnetic field, 
and $E_{\rm rot,i}$ is the initial rotational energy of an NS, 
\begin{eqnarray}\label{E_rot,i}
E_{\rm rot,i}\sim\frac{1}{2}I\left(\frac{2\pi}{P_{\rm i}}\right)^2. 
\end{eqnarray}
The strongest observed burst in FRB 121102 (burst number 2 in \cite{Mar+17}) 
gives the most stringent constraint on $L_{\rm sd}$. 
We use the burst with an isotropic luminosity of 
$L_{\rm FRB}\sim6\times10^{42}$ erg s$^{-1}$ \citep{Mar+17} detected 
at $\sim$4 yr after the first burst detection.

Second, the pulse width $w_{\rm p}$ has to be shorter than 
the rotational period $P$ for GRP-like emission. 
If the emission pattern is beamed, 
the pulse width relative to the rotation period should be 
short compared with the isotropic emission case.
Using the beaming fraction $f_{\Omega}$ for a circular cone geometry, 
the condition on the rotation period is  
\begin{eqnarray}\label{period_condition}
P>f^{-1/2}_{\Omega}w_{\rm p},
\end{eqnarray}
unless all three axes, magnetic axis, rotation axis, 
and line-of-sight are nearly aligned.
The temporal evolution of the rotation period is described by
\begin{eqnarray}\label{period}
P\sim P_{\rm i}\left(1+\frac{t}{t_{\rm sd}}\right)^{\frac{1}{n-1}}.
\end{eqnarray}
The observed Gaussian FWHM is $2.8-8.7$ ms for 17 bursts 
\citep{Spi+14, Spi+16, Sch+16}. 
Although the widest observed pulse width was $8.7$ ms 
for burst number 5 in \citet{Spi+16}, 
this pulse seems to be split into two components.
Here, we adopt the second widest width $8.0$ ms 
for burst number 10 in \citet{Spi+16}, 
though the following results would be almost comparable. 
Note that the width in the source rest frame is 
$(1+z)^{-1}$ times shorter than the observed one.

Figure \ref{limit} shows the constraints on the NS parameters $P_{\rm i}$ and $B_{\rm d}$
obtained from two conditions (\ref{luminosity_condition}) and (\ref{period_condition}). 
Each panel shows the case with different parameter sets ($f_{\Omega}$, $n$, $t$). 
Since the first burst of FRB 121102 were detected in 2012 \citep{Spi+14}, 
the NS age at which the strongest burst were detected \citep{Mar+17} 
is $t\gtrsim4/(1+z)~{\rm yr}\sim3.3$ yr in the source rest frame. 
Figure \ref{limit} (A) shows the fiducial case with isotropic emission ($f_{\Omega}=1$), 
dipole spin-down formula ($n=3$), and the lower bound of the NS age $t=3.3$ yr. 
Allowed ranges of parameters are very limited, 
$B_{\rm d}\sim10^{13}$ G and $P_{\rm i}\lesssim3$ ms. 
Such an NS with short $P_{\rm i}$ is very rare from the initial period distribution 
obtained by pulsar population studies \citep{FK06, Per+13}, 
the associated supernova remnants \citep{PT12}, 
and the total injected energy of pulsar wind nebulae (PWNe; \cite{TT13}).
The observed FRB luminosity is almost comparable to the spin-down luminosity 
in the allowed parameter region at $t=3.3$ yr. 
Then, the observed FRB flux density at 1.4GHz will rapidly decay toward $\lesssim 1$ Jy 
because the NS age in the allowed region is already $t\gg t_{\rm sd}$.

Figure \ref{limit} (B) shows the allowed parameter region 
in the anisotropic emission case, $f_{\Omega}=0.1$.
Since the intrinsic FRB luminosity is lower than the isotropic case, 
the allowed parameter region from the luminosity condition (\ref{luminosity_condition}) 
becomes larger than the fiducial case (figure \ref{limit} (A)).
The upper limit on the magnetic field is $B_{\rm d}\sim4\times10^{13}$ G. 
The fraction $w_{\rm p}/P$ should also reduce in the anisotropic emission case, 
so that the lower limit on the magnetic field becomes high 
at $t>t_{\rm sd}$ compared with the fiducial case.
As a result, the allowed range of the magnetic field is very limited, 
which is similar to the fiducial case.
The allowed range of the initial period becomes large, $P_{\rm i}\lesssim10$ ms.

The braking index $n$ is lower than that from the dipole spin-down formula ($n=3$) 
in most young pulsars (e.g., \cite{ELS17}). 
Figure \ref{limit} (C) shows the NS parameter region in the low braking index case, $n=2$.
In the region $t\ll t_{\rm sd}$, 
since conditions (\ref{luminosity_condition}) and (\ref{period_condition}) 
do not depend on $n$ (see equations \ref{L_sd} and \ref{period}), 
there is no allowed parameter region as the same as the case with $n=3$.
The allowed range of the initial period is $P_{\rm i}\lesssim2$ ms.
The allowed magnetic field depends on $P_{\rm i}$ and becomes low for short $P_{\rm i}$. 

The upper limit on the magnetic field $B_{\rm d,max}$ obtained from condition 
(\ref{luminosity_condition}) is 
$B_{\rm d,max}\propto(f_{\Omega}L_{\rm FRB})^{\frac{1-n}{4}}t^{-\frac{n+1}{4}}P_{\rm i}^{\frac{3-n}{2}}$.
On the other hand, the lower limit on the magnetic field $B_{\rm d,min}$ at $t\gg t_{\rm sd}$ 
derived from condition (\ref{period_condition}) is 
$B_{\rm d,min}\propto f_{\Omega}^{\frac{1-n}{4}}w_{\rm p}^{\frac{n-1}{2}}t^{-\frac{1}{2}}P_{\rm i}^{\frac{3-n}{2}}$.
From the two constraints on the dipole magnetic field, 
we obtain the constraint on the age of the NS $t$ in the source rest frame, described by
\begin{eqnarray}\label{age_constraint}
t&<&\frac{1}{2}IL_{\rm FRB}^{-1}\left(\frac{w_{\rm p}}{2\pi}\right)^{-2} \nonumber \\
&<&3.3\left(\frac{L_{\rm FRB}}{6\times10^{42}~{\rm erg~s}^{-1}}\right)^{-1}\left(\frac{w_{\rm p}(1+z)}{8~{\rm ms}}\right)^{-2}~{\rm yr}. \\
 & & \nonumber
\end{eqnarray}
Note that there is no allowed parameter region at $t\ll t_{\rm sd}$ 
if condition (\ref{age_constraint}) are not satisfied.
At $t=8/(1+z)~{\rm yr}\sim6.7$ yr, GRP-like emission model 
cannot explain the observations as shown in figure \ref{limit} (D).
The upper limit on the age (inequality \ref{age_constraint}) 
does not depend on the beaming fraction $f_{\Omega}$ and the braking index $n$.
Therefore, if the origin of FRB 121102 is the GRP-like emission from a very young NS, 
the NS was born in a short time before the first detection at 2012 
and FRB signal with flux density $\gtrsim1$ Jy will not be able to be detected after 2017. 

\section{DISCUSSION}
\label{sec:discussion}

We investigate the GRP-like emission model with circular cone emission geometry 
for an origin of repeating FRB 121102 
to give constraints on the NS parameters from the observations.
In addition to the luminosity condition that (a) the intrinsic luminosity of FRBs is 
lower than the spin-down luminosity as already discussed in some authors (e.g., \cite{L17}), 
we consider that (b) the intrinsic pulse width of FRBs is 
shorter than the rotational period. 
Condition (b) significantly reduces the allowed region of the NS parameters limited 
by only condition (a) as shown in Fig. \ref{limit}. 
Since the constraints are drawn from the properties of the burst only, 
the results do not depend on the properties of the persistent source and 
the circumburst environment discussed in other papers 
\citep{MBM17, L17, KM17, CYD17, DWY17}.
The determined parameters within narrow region are
the dipole magnetic field $B_{\rm d}\sim10^{13}$ G, 
and the initial period $P_{\rm i}\lesssim3$ ms, 
if the emission is isotropically emitted $(f_{\Omega}\sim1)$ 
and temporal evolution follows the dipole radiation formula $(n=3)$. 
The allowed dipole magnetic field becomes high for the anisotropic emission case ($f_{\Omega}<1$), 
and low for small braking index $(n<3)$, 
although the allowed range $\log(B_{\rm d,max}/B_{\rm d,min})$ does not depend on $f_{\Omega}$ and $n$, 
as shown in Figure \ref{limit}.
The maximum initial period becomes long for the anisotropic emission case, 
$P_{\rm i,max}\propto f_{\Omega}^{-1/2}$.

The FRB luminosity is comparable to the spin-down luminosity, 
$f_{\Omega}L_{\rm FRB}\sim L_{\rm sd}$, for the NS in the allowed region of Figure \ref{limit}. 
Although the radio efficiency required from the GRP-like model is higher than 
that of the observed Crab GRP $(\lesssim 0.01)$, such a high efficiency is seen 
in coherent magnetospheric radio emission from some old pulsars 
which reside close to the death line in $P$-$\dot{P}$ plane \citep{Sza+14}. 
A high efficiency state may change a spin-down behavior coincidence at FRBs. 
A suggestive switching behavior of the spin-down and radio emission has been 
reported from Galactic intermittent pulsars (e.g., \cite{Kramer+06}), 
which mechanism is proposed to be related with the changes of 
the dissipation rate of the electromagnetic energy in the magnetosphere 
(e.g., \cite{LST12}). 
Furthermore, the required efficiency is reduced by other parameters. 
For example, if an NS is significantly massive, 
the moment of inertia could be a factor of a few times larger than our adopted value. 
Then, the ratio $B_{\max}/B_{\min}$ could be high (from equation \ref{age_constraint}) 
and the required minimum efficiency could become an order of $\sim$0.1. 

Anisotropic emission case has been discussed by \citet{K16c}. 
In their extremely narrow, wandering beam model, 
the allowed parameter region for $P$ and $B_{\rm d}$ is much large. 
However, they have not considered the lower limit on the rotation period 
from the FRB pulse width. 
For the circular cone emission geometry as usually considered 
in the pulsar radio emission (e.g., \cite{R93}),
conditions (a) and (b) give significant constraints on 
the pulsar parameters (Figure \ref{limit}).
In addition, the constraint on the age (equation \ref{age_constraint}) 
does not depend on the beaming fraction $f_{\Omega}$.

The observed flux of repulsive bursts from FRB 121102 
has not shown apparent temporal evolution 
since the first detection \citep{Spi+14, Spi+16, Sch+16, Cha+17, Mar+17}. 
On the other hand, the spin-down luminosity is expected to already decay as
$L_{\rm sd}\propto t^{\frac{1+n}{1-n}}$ at $t\sim3.3$ yr 
(i.e., the allowed region in figure \ref{limit} satisfies at $t = 3.3~{\rm yr} > t_{\rm sd}$).
If the radio efficiency, $L_{\rm FRB}/L_{\rm sd}$, is constant, 
such behaviors are inconsistent. 
Note that the radio efficiency of GRP has been poorly understood \citep{EH16}.
The radio efficiency of normal pulse increases as the spin-down luminosity decreases
in Galactic pulsars whose ages are older than the spin-down timescale 
\citep{Sza+14}. 
If the radio efficiency of GRP (and FRB) has the similar to that of normal pulse, 
the flux may show no apparent temporal evolution until $t\sim3.3$ yr.
However, once the radio efficiency reaches to the maximum value $\sim1$ at the age
$t\sim3.3$ yr which does not depend on $f_{\Omega}$ and $n$ (inequality \ref{age_constraint}),
the observed radio flux density will rapidly decay ($\lesssim1$ Jy)
as the evolution of the spin-down luminosity even if the radio efficiency
keeps nearly maximum at $t > 3.3$ yr.  
If FRBs with flux $\gtrsim 1$ Jy will be detected 
from repeating FRB 121102 after 2017 (i.e., $t > 3.3$ yr), 
we need to consider an alternative model, such as the magnetically powered flare 
from a magnetar (e.g., \cite{PP07, L14, K16a}). 
Note that condition (\ref{period_condition}) is also working if the FRB emission occurs 
in the magnetar magnetosphere for the magnetically powered flare model.

Some other FRBs have comparable and wider observed pulse width 
(e.g., $w_{\rm p}\sim15.62$ ms for FRB 130729; \cite{Cham+16}).
However, after accounting for all instrumental and measurable propagation effects, 
the width of other all FRBs is smaller than $\sim3$ ms \citep{Sch+16}. 
Then, if other FRBs are also GRP-like emission and have similar luminosity 
$L\sim10^{42}-10^{43}$ erg s$^{-1}$, 
the activity timescale is $\lesssim30-300$ yr, much longer than that of FRB 121102. 

\subsection{Uncertainties of Pulse Width}

The observed pulse width would be broadened as results from multi-path propagation delay 
(e.g., \cite{LJ75, R77}).  
Then, the intrinsic pulse width of FRB would be overestimated. 
The scattering time in intergalactic medium is much shorter than the pulse width \citep{MK13}.
Using the scattering time-DM trend in Galactic pulsars \citep{LKMJ13, Cor+16}, 
the scattering time in the Galaxy is $\sim0.1-0.2$ ms, 
which is much shorter than the time-resolution of the measurements 
after de-dispersion \citep{Spi+14}. 
The contribution of the host galaxy to the scattering time is unknown. 
The detected angular broadening of the bursts and persistent radio source 
is similar to the expected our Galaxy scattering contribution \citep{Mar+17}. 
If the scattering time-DM trend in the host galaxy is also similar to that of the Galaxy, 
from the DM in host galaxy, 55 cm$^{-3}$ pc $\lesssim$ DM $\lesssim$ 255 cm$^{-3}$ pc \citep{Ten+17}, 
the scattering time is $10^{-3}-0.9$ ms. 
Observationally, no evidence of the scatter broadening was seen 
in all bursts of FRB 121102 \citep{Spi+14, Spi+16, Sch+16}. 
The upper limit on the observed scatter broadening for first burst is $<1.5$ ms \citep{Spi+14}.
The conservative upper limit on the scatter broadening is 
the minimum pulse width among the observed bursts, 
$2.8$ ms observed for burst number 6 in \citet{Spi+16}.
Even if we adopt the pulse width $w_{\rm p}=(8.0-2.8)/(1+z)$ ms $\sim 4.4$ ms 
as the upper limit on the rotational period, 
the upper limit on the age (inequality \ref{age_constraint}) is $t\lesssim7.8$ yr, 
only a factor of $\sim2$ larger than that in no scatter broadening case. 
Note that although the observed pulse width has broad range, 2.8-8.7 ms \citep{Spi+16}, 
this feature have also been seen in GRPs from the Crab pulsar (e.g., \cite{Mikami+16}).

The time resolution after de-dispersion of FRB 121102 pulses is $\sim1$ ms \citep{Spi+14, K16b}. 
Even if the light curves were analyzed in higher time resolution, 
the scattering time with $\sim1$ ms could be possible in current observational constraints \citep{Spi+14}.
Then, the pulsation structure could be washed out if the rotation period of the NS is $P\lesssim 1$ ms. 
The radio burst-like phenomena which continue up to $\sim10$ rotation periods have been known 
in some pulsars (e.g., \cite{LCM13}).
The minimum period of an NS is $\sim0.3-0.7$ ms which depends on the nuclear equation of state 
(e.g., \cite{HLZ99}). 
If the source of FRB 121102 is the sub-millisecond pulsar, 
the allowed range of the dipole magnetic field is $10^{11}$ G $\lesssim B_{\rm d}\lesssim10^{13}$ G 
for the age $t=3.3$ yr. 
Since the spin-down timescale becomes long for the NS with the low magnetic field 
and short rotation period, 
older NS age compared with the NS with long period ($P\gtrsim1$ ms) could be possible 
to explain the observations.
The upper limit on the age is $t\lesssim100$ yr for the pulse width $w_{\rm p}\sim1$ ms. 
The detail periodicity search in a burst (e.g., \cite{K16c}) 
and more stringent limit on the scatter broadening would be important  
to distinguish which range of period, $P\lesssim1$ ms or $P\gtrsim8$ ms, 
is appropriate as the source of FRB 121102. 

\subsection{Comparisons with Other Constraints}

Observed DM and its time-derivative could give significant constraints on 
the origin of the FRB (e.g., \cite{P16, MBM17, L17, KM17, CYD17}). 
Since an NS is formed as a result of core-collapse supernova, a young NS is 
generally surrounded by the dense ejecta.
For simplicity, a DM through the ejecta may be described by 
DM$_{\rm ej}\sim3f_{\rm ion}M_{\rm ej}/(8\pi v_{\rm ej}^2t^2m_{\rm p})$, 
where $M_{\rm ej}$ and $v_{\rm ej}$ are the mass and velocity of the ejecta 
which mainly consist of $\alpha$-elements, 
$f_{\rm ion}$ is the ionization fraction, and $m_{\rm p}$ is a proton mass. 
We assume singly ionization state for elements with the mean atomic mass number $\sim10$, 
so that $f_{\rm ion}\sim0.1$.
An NS which resides the allowed range in figure \ref{limit} has a huge initial rotation energy 
$E_{\rm rot,i}\sim3\times10^{52}(P_{\rm i}/1{\rm ms})^{-2}$ erg compared 
with a conventional supernova explosion energy ($\sim10^{51}$ erg). 
Since the age in the allowed region is $t\gg t_{\rm sd}$, 
the significant fraction of the rotation energy would be converted to the ejecta kinetic energy, 
$v_{\rm ej}\sim\sqrt{2E_{\rm rot,i}/M_{\rm ej}}$. 
Using the constraints $|d{\rm DM}_{\rm ej}/dt|\lesssim2$ cm$^{-3}$ pc yr$^{-1}$ 
for FRB 121102 estimated by \citet{P16}, 
the maximum ejecta mass is estimated as 
$(M_{\rm ej}/M_{\odot})\lesssim0.08(t/1{\rm yr})^{3/2}(P_{\rm i}/1{\rm ms})^{-1}$. 
Then, for GRP-like model, the NS would be formed with low mass ejecta 
such as ultra-stripped supernova 
($M_{\rm ej}\sim10^{-1}M_{\odot}$; e.g., \cite{KK14}), 
accretion-induced collapse ($M_{\rm ej}\sim10^{-3}M_{\odot}$; e.g., \cite{Des+06}), 
and binary NS merger ($M_{\rm ej}\sim10^{-4}-10^{-2}M_{\odot}$; e.g., \cite{Hot+13}), 
compared with conventional core-collapse supernova ($M_{\rm ej}\sim1-10M_{\odot}$). 

The persistent radio emission also gives the significant constraints on the NS age 
if the persistent source is the emission from the PWN (e.g., \cite{MKM16}). 
\citet{KM17} showed that for the ejecta mass $\sim0.1M_{\odot}$, 
the persistent source could be consistent with a PWN powered by an NS 
with our fiducial parameter sets, 
the dipole magnetic field $B_{\rm d}\sim10^{13}$ G, the initial rotation period $P_{\rm i}\sim 1$ ms, 
and the age $t\sim4$ yr. The range is consistent with our results from the observed FRB properties.

\begin{ack}
We are grateful to the anonymous referee for helpful comments.
We would like to thank Y. Ohira, S. J. Tanaka, T. Terasawa and R. Yamazaki for fruitful discussions. 
This work was supported by KAKENHI 16J06773 (S.K.), 15H00845, 16H02198 (T.E.) and 25400221 (S.S.).
\end{ack}


\begin{thebibliography}{}

\bibitem[Cao et al. (2017)]{CYD17}
Cao, X.-F., Yu, Y.-W., \& Dai, Z.-G. 2017, ApJL, 839, L20

\bibitem[Champion et al. (2016)]{Cham+16}
Champion, D. J., Petroff, E., Kramer, M., et al. 2016, MNRAS, 460, L30 

\bibitem[Chatterjee et al. (2017)]{Cha+17}
Chatterjee, S., Law, C. J., Wharton, R. S., et al. 2017, Nature, 541, 58 

\bibitem[Cheng et al. (1986)]{CHR86}
Cheng, K. S., Ho, C., \& Ruderman, M. 1986, ApJ, 300, 522

\bibitem[Cordes \& Wasserman (2016)]{CW16}
Cordes, J. M., \& Wasserman, I. 2016, MNRAS, 457, 232

\bibitem[Cordes et al. (2016)]{Cor+16}
Cordes, J. M., Wharton, R. S., Spitler, L. G., Chatterjee, S., \& Wasserman, I. 2016, arXiv:1605.05890

\bibitem[Dai et al. (2017)]{DWY17}
Dai, Z. G., Wang, J. S., \& Yu, Y. W. 2017, ApJL, 838, L7

\bibitem[Daugherty \& Harding (1982)]{DH82}
Daugherty, J. K., \& Harding, A. K. 1982, ApJ, 252, 337

\bibitem[Dessart et al. (2006)]{Des+06}
Dessart, L., Burrows, A., Ott, C. D., Livne, E., Yoon, S.-C., \& Langer, N. 2006, ApJ, 644, 1063

\bibitem[Eilek \& Hankins (2016)]{EH16}
Eilek, J. A., \& Hankins, T. H. 2016, JPlPh, 82, 635820302

\bibitem[Espinoza et al. (2017)]{ELS17}
Espinoza, C. M., Lyne, A. G., \& Stappers, B. W. 2017, MNRAS, 466, 147

\bibitem[Faucher-Gigu\'ere \& Kaspi (2006)]{FK06}
Faucher-Gigu\'ere, C.-A., \& Kaspi, V. M. 2006, ApJ, 643, 332

\bibitem[Goldreich \& Reisenegger (1992)]{GR92}
Goldreich, P., \& Reisenegger, A. 1992, ApJ, 395, 250

\bibitem[Haensel et al. (1999)]{HLZ99}
Haensel, P., Lasota, J. P., \& Zdunik, J. L. 1999, A\&A, 344, 151

\bibitem[Hotokezaka et al. (2013)]{Hot+13}
Hotokezaka, K., Kiuchi, K., Kyutoku, K., Okawa, H., Sekiguchi, Y., Shibata, M., \& Taniguchi, K. 2013, PhRvD, 87, 024001

\bibitem[Karuppusamy et al. (2010)]{KSv10}
Karuppusamy, R., Stappers, B. W., \& van Straten W. 2010, A\&A, 515, 36 

\bibitem[Kashiyama \& Murase (2017)]{KM17}
Kashiyama, K., \& Murase, K. 2017, ApJL, 839, L3

\bibitem[Katz (2016a)]{K16a}
Katz, J. I. 2016a, MPLA, 31, 1630013 

\bibitem[Katz (2016b)]{K16b}
Katz, J. I. 2016b, ApJ, 818, 19 

\bibitem[Katz (2017a)]{K16c}
Katz, J. I. 2017a, MNRAS, 467, L96

\bibitem[Katz (2017b)]{K17}
Katz, J. I. 2017b, MNRAS, 469, L39

\bibitem[Kirk et al. (2002)]{KSG02}
Kirk, J. G., Ski\ae raasen, O. \& Gallant, Y. A. 2002, A\&A, 388, L29 

\bibitem[Kleiser \& Kasen (2014)]{KK14}
Kleiser, I. K. W., \& Kasen, D. 2014, MNRAS, 438, 318 

\bibitem[Kramer et al. (2006)]{Kramer+06}
Kramer, M., Lyne, A. G., O'Brien, J. T., Jordan, C. A., \& Lorimer, D. R. 2006, Science, 312, 549

\bibitem[Lee \& Jokipii (1975)]{LJ75}
Lee, L. C., \& Jokipii, J. R. 1975, ApJ, 201, 532 

\bibitem[Li et al. (2012)]{LST12}
Li, J., Spitkovsky, A., \& Tchekhovskoy, A. 2012, ApJL, 746, L24

\bibitem[Lorimer et al. (2007)]{Lor+07}
Lorimer, D. R., Bailes, M., McLaughlin, M. A., Narkevic, D. J., \& Crawford, F. 2007, Science, 318, 777 

\bibitem[Lorimer et al. (2013a)]{LCM13}
Lorimer, D. R., Camilo, F., \& McLaughlin, M. A. 2013a, MNRAS, 434, 347 

\bibitem[Lorimer et al. (2013b)]{LKMJ13}
Lorimer, D. R., Karastergiou, A., McLaughlin, M. A., \& Johnston, S. 2013b, MNRAS, 436, L5 

\bibitem[Lyubarsky (2014)]{L14}
Lyubarsky, Y. 2014, MNRAS, 442, L9 

\bibitem[Lyutikov et al. (2016)]{LBP16}
Lyutikov, M., Burzawa, L. \& Popov. S. B., 2016, MNRAS, 462, 941

\bibitem[Lyutikov (2017)]{L17}
Lyutikov, M. 2017, ApJL, 838, L13

\bibitem[Macquart \& Koay (2013)]{MK13}
Macquart, J.-P., \& Koay, J. Y. 2013, ApJ, 776, 125 

\bibitem[Marcote et al. (2017)]{Mar+17}
Marcote, B., Paragi, Z., Hessels, J. W. T., et al. 2017, ApJL, 834, L8 

\bibitem[Manchester et al. (1985)]{MND85}
Manchester, R. N., Newton, L. M., \& Durdin, J. M. 1985, Nature, 313, 374 

\bibitem[Metzger et al. (2017)]{MBM17}
Metzger, B. D., Berger, E., \& Margalit, B. 2017, ApJ, 841, 14

\bibitem[Mikami et al. (2016)]{Mikami+16}
Mikami, R., Asano, K., Tanaka, S. J., et al. 2016, ApJ, 832, 212 

\bibitem[Michel (1988)]{M88}
Michel, F. C. 1988, Nature, 333, 644 

\bibitem[Murase et al. (2016)]{MKM16}
Murase, K., Kashiyama, K., \& M\'esz\'aros, P. 2016, MNRAS, 461, 1498 

\bibitem[Park et al. (2017)]{Park+16}
Park, S., Yang, J., Oonk, J. B. R., \& Paragi, Z. 2017, MNARS, 465, 3943 

\bibitem[Perera et al. (2013)]{Per+13}
Perera, B. B. P., McLaughlin, M. A., Cordes, J. M., Kerr, M., Burnett, T. H., \& Harding, A. K. 2013, ApJ, 776, 61 

\bibitem[Piro (2016)]{P16}
Piro, A. L. 2016, ApJL, 824, L32 

\bibitem[Popov \& Postnov (2010)]{PP07}
Popov, S. B., \& Postnov, K. A. 2010, 
in Proc. of Conf. Dedicated to Viktor Ambartsumian's 100th Anniversary, 
Evolution of Cosmic Objects Through Their Physical Activity, Vol. 129, ed. H. A. Harutyunian, 
A. M. Mickaelian, \& Y. Terzian (Yerevan: NASRA), 129

\bibitem[Popov \& Turolla (2012)]{PT12}
Popov, S. B., \& Turolla, R. 2012, Ap\&SS, 341, 457 

\bibitem[Rankin (1993)]{R93}
Rankin, J. 1993, ApJ, 405, 285 

\bibitem[Rickett (1977)]{R77}
Rickett, B. J. 1977, ARA\&A, 15, 479 

\bibitem[Spitler et al. (2014)]{Spi+14}
Spitler, L. G., Cordes, J. M., Hessels, J. W. T., et al. 2014, ApJ, 790, 101 

\bibitem[Spitler et al. (2016)]{Spi+16}
Spitler, L. G., Scholz, P., Hessels, J. W. T., et al. 2016, Nature, 531, 202 

\bibitem[Scholz et al. (2016)]{Sch+16}
Scholz, P., Spitler, L. G., Hessels, J. W. T., et al. 2016, ApJ, 833, 177 

\bibitem[Szary et al. (2014)]{Sza+14}
Szary, A., Zhang, B., Melikidze, G. I., Gil, J., \& Xu, R.-X. 2014, ApJ, 784, 59 

\bibitem[Tanaka \& Takahara (2013)]{TT13}
Tanaka, S. J., \& Takahara, F. 2013, MNRAS, 429, 2945 

\bibitem[Tendulkar et al. (2017)]{Ten+17}
Tendulkar, S. P., Bassa, C. G., Cordes, J. M., et al. 2017, ApJL, 834, L7 

\bibitem[Tendulkar et al. (2016)]{TKP16}
Tendulkar, S. P., Kaspi, V. M., \& Patel, C. 2016, ApJ, 827, 59 

\bibitem[Thornton et al. (2013)]{Tho+13}
Thornton, D., Stappers, B., Bailes, M., et al. 2013, Science, 341, 53 

\end{thebibliography}
\end{document}